\documentclass[12pt]{article}
\usepackage{amsmath,amssymb,amsthm}
\usepackage{latexsym,graphicx,color,subfigure,mathrsfs}
\usepackage{natbib}
\usepackage{epstopdf}

\def\vl{v_{\rm{los}}}
\def\mm{M_{\rm{min}}}
\def\rt{r_{\rm{tidal}}}
\def\Tr{{\rm Tr}}

\def\p{\partial}

\def\D{\mathcal{D}}

\def\hsp{\, ,\hspace{.7cm}}
\def\=:{=\hspace{-.7em}\raisebox{1.1ex}{.}\hspace{.1em}\raisebox{-0.2ex}{.} }

\newcommand {\beq}{\begin{eqnarray}}
\newcommand {\eeq}{\end{eqnarray}}

\def\diag{{\rm diag}}

\theoremstyle{definition}

\makeatletter
\@addtoreset{equation}{section}

\renewcommand{\thefootnote}{\fnsymbol{footnote}}
\makeatother

\makeatletter
\newcommand{\thetablename}{Table}
\def\fnum@table{\thetablename\ \thetable}
\makeatother

\begin{document}

\thispagestyle{empty}

\begin{center}
{\Large \bf Binary Satellite Galaxies}  
\\[25mm]
{\large Jarah Evslin}\footnote{jarah(at)ihep.ac.cn}
\vskip 6 mm

\bigskip
{\it
TPCSF, Institute of High Energy Physics, CAS\\
YuQuanLu 19B,
Beijing 100049, P.R.~China\\
}

\bigskip

\bigskip

\bigskip

\bigskip
\bigskip
\bigskip

{\bf Abstract}\\[5mm]
{\parbox{15cm}{\hspace{4mm}
%\small

\noindent
Suggestions have appeared in the literature that the following five pairs of Milky Way and Andromeda satellite galaxies are gravitationally bound: Draco and Ursa Minor, Leo IV and V,   Andromeda I and III,  NGC 147 and 185, and the Magellanic clouds.    Under the assumption that a given pair is gravitationally bound, the Virial theorem provides an estimate of its total mass and so its instantaneous tidal radius.   For all of these pairs except for the Magellanic clouds the resulting total mass is 2 to 4 orders of magnitude higher than that within the half light radius.   Furthermore in the case of each pair except for Leo IV and Leo V,  the estimated tidal radius is inferior to the separation between the two satellites.  Therefore all or almost all of these systems are not gravitationally bound.   We note several possible explanations for the proximities and similar radial velocities of the satellites in each pair, for example they may have condensed from the same infalling structure or they may be bound by a nongravitational interaction between dark matter and an extremely light particle.%three possible explanations for this fact.  First, the pairs were never bound.  Second, each binary system was bound or condensed from the same bound structure but is passing near its host for the first time and is in the process of being disrupted.  Third, the pairs of dark matter halos are bound by a long-distance, non-gravitational force caused by an interaction between dark matter and a particle lighter than $10^{-25}$~eV.

}}
\end{center}
\newpage
\pagenumbering{arabic}
\setcounter{page}{1}
\setcounter{footnote}{0}
\renewcommand{\thefootnote}{\arabic{footnote}}
%%%%%%%%%%%%%%%%%%%%%%%%%%%%%%%%

\section{Introduction}

The dwarf spheroidal satellite galaxies in orbit about the Milky Way and the Andromeda Galaxy are the purest known accumulations of dark matter.  Thus they provide natural laboratories for understanding the properties of dark matter without the complexities of baryonic physics.   However this very advantage is also their limitation.  Dwarf spheroidal galaxies contain essentially no gas.  The only tracers available to determine the profiles of their dark matter halos are stars.  However these stars are dispersion supported and a degeneracy \citep{binney} prevents the Jeans equation from uniquely determining the underlying dark matter distribution in terms of the projected stellar dispersion.  Instead, for each chemically distinct population of stars \citep{chimica} the stellar dispersion produces only a single number, the mass contained within the projected half light radius \citep{walker,wolf}.  

One obvious limitation of stellar tracers is that, as a result of Birkhoff's theorem \citep{jebsen}, they only yield information about the dark matter halo in the region inhabited by stars.  If the stars do not extend out to the far reaches of the halo then even the size of the halo is unknown.  This is the case for the dwarf galaxies in our local group, so far tracers have produced no convincing evidence for an edge of a halo, in fact there is no clear reduction in the stellar dispersion at the greatest radii at which it has been measured.

In this note we observe that in the case of binary satellite galaxies more information is available.  15 years ago, \citet{ngc} suggested that Andromeda's satellite galaxies NGC 147 and 185 may be gravitationally bound to each other and used this assertion to estimate their masses.  Below we extend this argument to systematically estimate the total masses of five candidate binary pairs that have appeared in the literature, exclusively relying upon the local group member data assembled in \citet{elenco}.  From the estimated masses we can compare the gravitational attraction between the two satellites in a pair to the tidal force exerted by their host.  We will find that in essentially every case the tidal force wins, the only exception being Leo IV and Leo V if at least  99.9\% of the mass in their halos lie beyond the region in which their stars have been identified.  We conclude that it is unlikely that any of these pairs is gravitationally bound.

\section{Mass Estimates}\label{masssez}

\subsection{Methodology} \label{logica}

To estimate the total masses of the binary systems under the assumption that they are gravitationally bound, we rely upon Newtonian gravity.  Let the two galaxies have masses $M_1$ and $M_2$ and speeds $v_1$ and $v_2$.  In the center of mass frame
\beq
M_1v_1=M_2v_2
\eeq
and so
\beq
v_1=\frac{M_2}{M_T}v\hsp
v_2=\frac{M_1}{M_T}v
\eeq
where the relative velocity and total mass are
\beq
v=v_1+v_2\hsp
M_T=M_1+M_2.
\eeq
Now the total kinetic and potential energies are
\beq
T=\frac{1}{2}M_1 v_1^2+\frac{1}{2}M_2 v_2^2=\frac{v^2}{2}\frac{M_1M_2}{M_T}\hsp
U=-\frac{G_N M_1 M_2}{d}
\eeq
where $G_N$ is Newton's constant and $d$ is the distance separating the two galaxies.  In principle $U$ also contains a positive term \citep{hubble} which incorporates the fact that the expansion of the universe tends to separate the two galaxies.  Such a term would increase the masses that we will derive below.  However for the small separations in the binary systems to which we will apply this formula, the correction is insignificant as compared with our other sources of error.

If we assume that the binary system is gravitationally bound then the total energy is negative and so
\beq
M_T > \frac{v^2 d}{2G_N}.
\eeq
Generally only the line of sight velocity is available.  Let $\vl$ be the difference between the line of sight velocities of the two galaxies in a pair. As $v_l\leq v$ one may write a weaker inequality
\beq
M_T > \frac{v^2 d}{2G_N}\geq\mm=\frac{\vl^2 d}{2G_N}.
\eeq

Alternately, instead of a lower bound $\mm$\ on the total mass, one may be interested in an estimate.  Assuming isotropy, a rough estimate for $v$ is
\beq
v^2=3\vl^2.
\eeq
The total mass can be estimated using the Virial theorem, assuming the system to be gravitationally bound, in equilibrium and in a roughly average configuration and approximating the expectation value of the inverse radius to be the reciprocal of the expectation value of the radius.  In an average configuration $2T=-U$ and so we arrive at our final formula for the estimate of the total mass of the binary system
\beq
M_T = \frac{v^2 d}{G_N} = \frac{3\vl^2 d}{G_N}. \label{m}
\eeq
Note that this approach differs from that of \citet{davis} in which only the projected distance is used as we do not assume that the vector separating the two galaxies is parallel to their relative velocity.  The catalog \citet{elenco} provides the full 3-dimensional separation $d$ with sufficient precision for (\ref{m}) to yield a useful estimate.

\subsection{Calculations}

In this section we will apply Eq.~(\ref{m}) to estimate the total masses of the 5 pairs of local group satellite galaxies which have been claimed to be gravitationally bound in the literature.   We will first consider the Magellanic clouds, whose gravitational association has long been suspected.  The line of sight velocities have been measured precisely \citep{mclos} and are respectively $262.2$ km/s and $145.6$ km/s for the LMC and the SMC.  The angle between these lines of sight is $20.7^\circ$ and so they are rather far from parallel.  Therefore simply subtracting the two line of sight velocities to obtain a relative line of sight velocity is a poor approximation.  However, unlike the other binary systems considered here, the tangential velocities of both the LMC \citep{lmc} and SMC \citep{smc} have been measured by the Hubble space telescope.    

In the (east, north, radial) basis the best fit velocities of the clouds with respect to the Sun, in km/s, are
\beq
v^{(spherical)}_{\rm{LMC}}=(482,104,262)   \hsp
v^{(spherical)}_{\rm{SMC}}=(340,-341,146).
\eeq
Let $\theta$ and $\phi$ be the spherical coordinate angles corresponding to the right ascension and declination respectively
\beq
(\theta,\phi)_{\rm{LMC}}=(-69.8^\circ,80.9^\circ)\hsp
(\theta,\phi)_{\rm{SMC}}=(-72.8^\circ,13.2^\circ).
\eeq
Then the velocities in Cartesian coordinates are easily found, in units of km/s, to be
\beq
v^{(Cartesian)}_{\rm{LMC}}=(-446,262,-210)   \hsp
v^{(Cartesian)}_{\rm{SMC}}=(-353,267,-240).
\eeq
The norm of the best fit relative velocity is
\beq
v=|v^{(Cartesian)}_{\rm{LMC}}-v^{(Cartesian)}_{\rm{SMC}}|=98\rm{\ km/s}.
\eeq

As we have the absolute velocity difference and not just the line of sight velocity difference, one may determine the mass using the first equality in Eq.~(\ref{m}).  We obtain
\beq
M_T=5.5\times 10^{10}M_{\odot}\hsp
\mm=M_T/2=2.7\times 10^{10}M_{\odot}
\eeq
where as described above the minimum mass $\mm$ is calculated by setting the sum of the kinetic and potential energy to zero.  These results are summarized in Table.~\ref{galtab}.

\begin{table}[position specifier]
\centering
\begin{tabular}{c|l|l|l|l|l|l}
&Draco\&UMi &Leo IV\&V&And I\&III&NGC 147\&185&LMC\&SMC\\
\hline\hline
$d$ (kpc)&$23^{+2}_{-0}$&$25^{+11}_{-10}$&$33^{+16}_{-0}$&$59^{+39}_{-36}$&$24^{+2}_{-1}$\\
\hline
angular sep.&$17.4^\circ$&$2.8^\circ$&$2.5^\circ$&$1.0^\circ$&$20.7^\circ$\\
\hline
$\vl$ (km/s)&$44.1\pm 0.1$&$41.0\pm 3.4$&$30.2\pm 2.3$&$10.7\pm 1.4$&$v=98$\\
\hline
$M(r<r_{\rm{h-l}})$ ($10^{6}M_\odot$)&$10.2$&$1.2$&$25$&$121$&\\
\hline
$M_T$ ($10^{9}M_\odot$)&$31$&$30$&$21$&$5$&$55$\\
\hline
$\mm$\  ($10^{9}M_\odot$)&$5.0\pm 1.0  $&$4.9\pm 2.8$&$3.5^{+3.6}_{-1.2}$&$0.80\pm 0.62$&$28$\\
\hline
$\rt$\ (kpc)&$13$&$28$&$8$&$12$&$19$\\
%\hline
%&$ $&$ $&$ $&$ $&$ $\\
\hline
\hline
\end{tabular}
\caption{Mass estimates and tidal radii of 5 candidate binary systems in our local group, calculated under the assumption that these pairs are gravitationally bound}
\label{galtab}
\end{table}

In the cases of the other pairs of galaxies, transverse velocities are not available and so the estimation of the mass is simpler.  We simply subtract the velocities relative to the Sun, taken from \citet{elenco}, to obtain $\vl$ which we insert into Eq.~(\ref{m}), yielding $M_T$.  As described above, dividing this by 6 we find the lower bound $\mm$.  Of course the lines of sight are never precisely parallel and so this naive subtraction will always overestimate the relative velocity.  However the error introduced using this crude approximation is much smaller than the error in the approximation that the total velocity squared is three times the line of sight velocity squared, and so we will ignore it.  In fact the later error is so large that we will never include error bars in our estimations of $M_T$.

The next pair of galaxies that we will consider is NGC 147 and 185, which were proposed to be gravitationally bound in \citet{ngc}.  The author used their projected separation together with the assumption that they are gravitationally bound to obtain a lower bound on the total mass of the binary system using the relation given in \citet{davis}.   The author obtained a best fit of $\mm=2.7\times 10^8M_\odot$.  However in the case of this binary system the radial separation is about three times larger than the projected separation, and so using Eq.~(\ref{m}) the 3-dimensional separation gives an appreciably tighter constraint on the total mass.  Combined with the fact that the relative velocity reported in \citet{elenco} is about 15\% greater than that used by \citet{ngc}, we find a minimum mass of $\mm=(8\pm 6)\times 10^8M_\odot$ and a total mass 6 times greater, as is summarized in Table~\ref{galtab}.

Chronologically the next suspected bound pair of satellite galaxies consists of Leo IV and Leo V.  The association of this pair has been suspected since the discovery of Leo V by \citet{leo5}, who estimated the probability of their close association being by chance to be less than 1\%.  However \citet{leo5} attribute their proximity in position and velocity to their cohabitation in the same stream and do not claim that Leo IV and V are presently gravitationally bound to each other.   Recently the stream proposal has been weakened by the observation \citep{nessunponte} that the stellar density observed between Leo IV and V is a foreground stream, more than 100 kpc from the galaxies themselves.   

On the other hand, \citet{leo1,leo2} have suggested that this pair is indeed bound and used this assumption to determine a lower bound on their masses, reporting $\mm=(8\pm 4)\times 10^9M_\odot$ and $\mm=(3.5\pm 1.9)\times 10^{10}M_\odot$ respectively.  The first of these estimates uses the assumption \citep{davis} that the relative velocity of the two galaxies is parallel to the line separating the two galaxies, and so it is not truly a lower bound.   While \citet{leo2} confirmed the results of \citet{leo1} using the same methodology, they obtain a number of different estimates using different methods.  For example, by asserting that Leo IV and V will not be separated by tidal forces from the Milky Way they conclude that the minimum mass must be $\mm=4\times 10^{10}M_\odot$.  Such estimates will be the subject of Sec.~\ref{maree} and so we will not comment on them further here.  Using the logic described in Subsec.~\ref{logica}, a separation of $d=25^{+11}_{-10}$ kpc and a relative line of sight velocity of $\vl=41.0\pm 3.4$ km/s has led us to obtain a minimum total mass of $\mm=(4.9\pm 2.8)\times 10^9 M_\odot$, which is lower than that of \citet{leo1} since we do not assume the relative velocity and displacement of the galaxies to be parallel.  Using Eq.~(\ref{m}) we obtain an expected total mass of $M_T=3\times 10^{10}M_\odot$.

Recently two more candidate binary satellite galaxy systems have been identified by \citet{fattahi}.  One pair consists of Andromeda's dwarf spheroidal satellites And I and And III.   Like NGC 147 and 185, being satellites of Andromeda, the lines of sight are nearly colinear, with a difference of only $2.5^\circ$, and so $\vl$ can be estimated reliably as the difference between the two line of sight velocities, yielding $\vl=30.2\pm 2.3$ km/s.  Due to the relatively high luminosities of these galaxies, their line of sight distances are well estimated and so the true 3-dimensional distance between these galaxies can be determined to be $d=33^{+16}_{-0}$ kpc where the asymmetry in the errors is a result of the fact that the tangential displacement places a solid minimum on the distance between the galaxies.   The resulting halo mass estimate is then $M_T=2.1\times 10^{10}M_\odot.$  Surprisingly, despite the fact that this system contains orders of magnitude more stars than the ultrafaint pair Leo IV and Leo V, the estimated halo mass is marginally lower.   In fact the correlation between luminosity and halo mass in Table~\ref{galtab} is quite weak, with only a slightly higher mass for the Magellanic clouds than for the dwarf spheroidal galaxies.

The last pair proposed by \citet{fattahi} consists of the Draco and Ursa Minor dwarf spheroidal Milky Way satellites.  These are only 76 kpc away.  This is a disadvantage as it means that, due to their 23 kpc separation the lines of sight are separated by $17.4^\circ$ and we may therefore expect our approach to somewhat overestimate the total mass.  However it also means that the upcoming Gaia mission may well be able to determine the transverse velocities of the stars, whose applications will be described below.   For now, we will rely upon the line of sight velocities and we will simply subtract them to arrive at a relative line of sight velocity of $\vl=44.1\pm 0.1$ km/s.  Inserting the relative separation of $d=23^{+2}_{-0}$ kpc into Eq.~(\ref{m}) then leads to a mass estimate of $M_T=3.1\times 10^{10}M_\odot$, essentially equal to that of the much fainter pair Leo IV and V.

\section{Tidal Radii} \label{maree}

The total masses obtained in the previous section are quite imprecise.  The Virial theorem inspired estimate that the kinetic energy is half of the negative potential energy already introduces a potential factor of two, and the estimate that the relative velocity squared is three times the line of sight velocity squared introduces a potential factor of three.  By comparison the geometric approximations are mild.  Nonetheless, the order of magnitude of the results is a robust consequence of the tenuous assumption that these binary systems are gravitationally bound.  Not only is it robust, but it is nontrivial as it generally exceeds the mass $M(r<r_{\rm{h-l}})$ deduced from stellar tracers using the formulae of \citet{walker,wolf} by three orders of magnitude.   

Nonetheless these two mass estimates are in principle mutually consistent, the tracers suggest that the inner 300 pc contain about $10^7M_\odot$ \citep{strigari} while one can see from Table~\ref{galtab} that the later suggests of order $3\times 10^{10}M_\odot$ within a radius of about 10 kpc.   While in principle these two estimates may be mutually consistent, in practice the condition that a dark matter profile must satisfy both of these conditions is quite powerful, as it implies that the density falls off on average at most as $1/r$ in this regime, if not more slowly.  This would be a challenge for a pseudoisothermal halo, since it would require that the constant density core extend far beyond the furthest identified stars.  In the case of an NFW profile it would require the $1/r$ behavior to continue to about 10 kpc.  For now such a discussion is pure speculation, but the measurement of tangential velocity dispersions in the near future in some of these systems can test and distinguish these somewhat extreme scenarios.

In this section we will instead consider a more concrete calculation.  As noted by \citet{leo2} in this context, in order for a binary satellite galaxy system to be gravitationally bound it is not sufficient that the sum of the kinetic and potential energy be negative.  It is also necessary that the binding be sufficiently strong so as not to be disrupted by tidal forces arising from the host galaxy.  This can be restated simply as the condition that the tidal radius for each galaxy be greater than the separation between the two galaxies in the pair.  This condition is a consequence of Newtonian gravity, and we will see that it fails for most of the candidate pairs of galaxies in our sample.  

The most straightforward formula for the tidal radius $r_t$ of a mass $M$ gravitationally bound many body system in equilibrium subjected to a tidal force from a body of a mass $M_g$ at a distance $R$ is the instantaneous tidal radius of \citet{tidal1}
\beq
r_t=R\left(\frac{M}{2M_g}\right)^{1/3}.
\eeq
However in practice the system in question will be in orbit about the massive body and so one must also consider the contribution of the centrifugal force.  For a circular orbit this leads to a tidal radius of \citep{tidal2}
\beq
r_t=R\left(\frac{M}{3M_g}\right)^{1/3}. \label{tidal}
\eeq
In principle the tidal radius is further reduced as a consequence of the fact that orbits are generally elliptical and so the bound system will eventually pass closer to the massive body, and so the true tidal radius should be evaluated at the perigalactic point.  However, as has been noted by \citet{tidal2}, due to Birkhoff's theorem at smaller radii the system feels less gravity from the massive body which leads to a decrease in the effective $M_g$ at lower radii, thus increasing the tidal radius.  To determine the first effect one must know the ellipticity of the orbit, which for now is only available for the Magellanic clouds but may be estimated by Gaia in the case of Ursa Minor and Draco and perhaps even Leo IV and V.  The second effect can be incorporated by using a mass model of the host galaxy.  We leave both improvements to future work, now simply using (\ref{tidal}) and making the poor assumption that the Milky Way and Andromeda are point masses of mass $10^{12}M_\odot$ and $2\times 10^{12}M_\odot$ respectively.  The poorness of these approximations is somewhat alleviated by the fact that only the cuberoots of the masses appear in Eq.~(\ref{tidal}).

Approximating the mass of each satellite galaxy in a binary system to be one half of the total mass $M_T$ in Table~\ref{galtab} and setting $R$ equal to the distance between the center of mass of the binary systems and their hosts, we use Eq.~(\ref{tidal}) to produce tidal radii $\rt$ for the 10 dwarfs in the 5 binary systems, reporting the results in Table~\ref{galtab}.  Note that due to our equal mass approximations, the tidal radii of the two galaxies in a pair are equal.   The necessary condition for a pair of galaxies to remain bound despite the tidal force is then simply $r_t>d$.  Since these are two body systems and not spherically symmetric globular clusters as were considered by \citet{tidal1,tidal2} one may object that the two satellite galaxies may orbit each other on a plane which is just by chance perpendicular to the line between the binary system and the host galaxy, in which case there would be no tidal force and so the condition $r_t>d$ is not really a necessary condition for the binary system to be gravitationally bound.  However such a circumstance cannot persist throughout an entire orbit about the host galaxy, and for binaries on the first pass about their host it is anyway difficult to determine if the pair has already been disrupted.  Thus the existence of these rare geometric configurations will not appreciably affect our conclusions.

Our conclusion, as is evident from Table~\ref{galtab}, is that in general the condition $r_t>d$ is not satisfied.  It is only satisfied in the case of Leo IV and V and even in this case $r_t$ and $d$ are almost equal.  %In several cases it is only excluded by about 1$\sigma$.  The strongest exclusion arises in the case of And I and III, although an inclusion of the asymmetric errors in NGC 147 and 185 would also reveal a fairly robust exclusion.  Thus while statistical fluctuations may allow some of these systems to be stable, it is highly likely that the condition $r_t>d$ is not satisfied for a majority of the 5 candidate binary systems.

For \citet{leo2} the fact that this condition fails simply implied that one needs to impose a stronger lower bound on the mass of the galaxies such that the attraction can overcome the tidal force.  However in our study we have not only provided a lower bound for the galactic masses, but we have also provided an average value based on isotropy and the Virial theorem.  As it is the cube root of the mass which enters in the formula for the tidal radius, these approximations would need to be much worse than one would expect statistically in order for the gravitational attraction of these systems to overcome the tidal forces of their hosts.  As we have determined an average value for the masses and not simply a lower bound, the tidal radii that we here derive are also average values and not lower bounds, and thus it is difficult to evade the conclusion that in most of these systems the tidal radii are less than the separations and so one expects the tidal forces to win.  Thus we conclude that most or all of the pairs considered here are not gravitationally bound.

\section{Comparison with Millennium Simulations} \label{finsez}

In this note we have used the assumption that a set of 5 pairs of satellite galaxies are gravitationally bound to calculate some of their characteristics.  One straightforward conclusion of this study is that most or all of these systems are simply not gravitationally bound.  However the existence of these five pairs is already interesting.  As one can see in Table \ref{galtab}, all 5 of these pairs are extremely close both in position space and in line of sight velocity space.   The two galaxies in each binary system appear to be separated by less than 60 kpc in each case and the line of sight velocities differ by less than 45 km/s, with a 98 km/s relative 3-dimensional velocity for the Magellanic clouds.  

This can be compared with the pairs identified using the Millennium simulation \citep{millennium} by \citet{millenniumstudia}.   In the left panel of Fig. 5 of that study, more than a thousand blue dots represent various pairs of satellite halos separated by distances of less than about 350 kpc.  In this figure one can observe both the spatial and the velocity separations of the galaxies in the pairs, and one can see that not a single pair in this simulation is as close in phase space as any of the local group pairs that have been identified in the literature and summarized in our Table \ref{galtab}.   This extends the observation of \citet{fattahi} that simulations tend to produce less pairs of satellites than have been identified in our local group.  Notice that this observation is independent of whether the pairs are gravitationally bound or not, it is simply a consequence of the distributions of the distances between the pairs and their relative velocities.

\citet{millenniumstudia} restrict their attention to pairs with masses above about $10^9\ M_\odot$ and so one may object that the Magellanic clouds are the only pair considered here which fits their criteria.   These are separated by 24 kpc and have a relative 3-dimensional velocity of 98 km/s.  However an inspection of the left panel of Fig. 5 of \citet{millenniumstudia} indicates that no pair in their study had both a smaller or equal separation and a smaller or equal relative velocity than the Magellanic clouds.  This already indicates that the satellite pairs in our local group are quite different from those found using the Millennium simulation, at least using the merging history assumed in that study.  

\section{Pairs from Mergers with a Common Progenitor}

If these pairs are not gravitationally bound, why are they so close physically and why do they have similar radial velocities and luminosities?

\subsection{Scenarios}

A first guess may be that these binary systems simply were never gravitationally bound.  Perhaps it is a shear coincidence that the two members of a pair have similar positions, velocities and luminosities.  For individual pairs the probability of such an occurrence has been estimated by \citet{leo5,fattahi} and, considering the number of pairs, it is well below 1\%.  Considering the ubiquitous existence of structures in the phase space distribution of the Milky Way's \citep{milkydisco} and Andromeda's \citep{andromedadisco} satellite galaxies as well as the filamentary structure around NGC3109 \citep{filament} this possibility seems quite unlikely.  

\citet{belreview} has suggested that the phase space correlations in the Milky Way and Andromeda's satellite systems could be explained if these satellites condensed from a once gravitationally bound object, or a piece of such an object, which has been accreted by the host galaxy.   We will now argue that such a scenario may also be able to explain the abundance of gravitationally unbound satellite pairs, however it nonetheless requires a rather restrictive accretion history which may motivate the search for alternative explanations.

Such scenarios can be divided into two categories.  First it may be that both galaxies in a given pair were part of an extended structure which merged with our local group.  This was essentially proposed by \citet{leo5} for Leo IV and V, although the proposed structure was later revealed to be a foreground.  The second possibility is that these pairs existed as bound binaries but are now approaching their host galaxies for the first time and so are in the process or disassociating.   In the case of Milky Way satellites, as data arrives concerning tangential velocities of these systems a more accurate picture of their past orbital histories will emerge and these scenarios may be evaluated.  

In this section we will see that fairly strong assumptions are necessary in both cases.  The first category may require recent large mergers in both the Milky Way and the Andromeda systems or else it is difficult to see why the pairs should be separating just now.  Similarly the second may require us to live at a special moment when all of these pairs are arriving close to their host for their first time.  

\subsection{Comparison of Energy and Angular Momenta}

How can such scenarios be tested?

First one must determine just when the satellite galaxies in each pair formed.  If indeed they are not gravitationally bound to each other, then the estimates of their masses in Sec.~\ref{masssez} are unmotivated.  The masses must still satisfy the lower bounds of order $10^7\ M_{\odot}$ given the dispersions of their stars, but can well be much less than $10^{10}\ M_{\odot}$ so as to agree with the results of simulations.  As a result of these low masses, at the distances of 10 kpc or more by which these pairs are separated, the gravitational attraction between the galaxies in a given pair is irrelevant.  

Thus each galaxy in a pair follows an independent orbit about the host galaxy.   We know that the galaxies in each pair are separated by about 30 kpc and have relative velocities of order 30 km/s.   Thus one might suspect that they separate quickly and so such pairs should not exist for long.   However it could be that, as a result for example of the compactness of their common progenitor, the two satellites in a pair have essentially the same center of mass energy and angular momentum\footnote{In this discussion we are actually interested in the total energy and angular momentum per unit mass of the satellite, but for brevity we will omit the phrase, ``per unit mass".} about their host.  In this case they would inhabit distinct orbits with the same ellipticity and perigalactic distance and so, while the distance between the satellites and the difference between their radial velocities would change in time, this change would be periodic and so such a small difference could be stable over the cosmological time since these satellites formed.

Is it possible that the satellites in each pair indeed have the same total energy and angular momentum about their host?  Consider two Milky Way satellites which are separated from the Milky Way by distances $r_1$ and $r_2$ with radial velocities $v^r_1$ and $v^r_2$.  As we are much closer to the center of the Milky Way than the satellites, these radial velocities with respect to the Milky Way can be well estimated by simply correcting the radial velocity with respect to the Sun by the Sun's motion about the center of the Galaxy.  This would not be the case for satellites of Andromeda.  Let $v^t_1$ and $v^t_2$ be the magnitudes of their tangential velocities, in other words the norm of the velocity 2-vector normal to the radial direction from the center of the Milky Way to the satellite.  Now let $M$ be the mass of the Milky Way out to the distance $r_1$.  Since $r_1$ and $r_2$ are close, we will make the further approximation that this is equal to the mass of the Milky Way out to $r_2$.

Now the condition that both satellites in a pair have the same angular momentum is
\beq
r_1 v^t_1=r_2 v^t_2
\eeq
whereas the condition that they have the same center of mass kinetic plus potential energy is
\beq
\frac{1}{2}(v^r_1)^2+\frac{1}{2}(v^t_1)^2-\frac{GM}{r_1}=\frac{1}{2}(v^r_2)^2+\frac{1}{2}(v^t_2)^2-\frac{GM}{r_2}.
\eeq
Combining these conditions we can find the tangential velocity squared of either satellite
\beq
(v^t_1)^2=2GM\frac{r_2}{r_1(r_1+r_2)}+\left[(v^r_2)^2-(v^r_1)^2\right]\frac{r_2^2}{r_2^2-r_1^2}.
\eeq
To leading order in an expansion about $r=r_1$ with respect to $(r_2-r_1)/r_1$ this reduces to
\beq
(v^t_1)^2=\frac{GM}{r}+\left[(v^r_2)^2-(v^r_1)^2\right]\frac{r}{2(r_2-r_1)}.
\eeq
We may recognize the first term on the right hand side as $v^2$ for a circular orbit, where $v$ should be 220 km/s at small distances from the Milky Way and then eventually drop to zero.   

\subsection{Tangential Velocities of Milky Way Satellites}

What would this imply for our three Milky Way satellite pairs?

Let us begin with Draco and Ursa Minor.  The radial velocities are known quite well.  In the case of Draco and Ursa Minor they are respectively $v_1=-96$ km/s and $v_2=-85$ km/s.  In particular, Draco is infalling faster than Ursa Minor.  This leads us to a tangential velocity for Draco of 
\beq
(v^t_1)^2=40,000 {\rm{(km/s)}}^2+2,000 {\rm{(km/s)}}^2\frac{38 {\rm{\ kpc}}}{(r_2-r_1)}. \label{dravel}
\eeq
The radial distances are known somewhat less precisely $r_1=76\pm 6$ kpc and $r_2=78\pm 3$ kpc.  Therefore it is not known which is closer.  

\begin{table}[position specifier]
\centering
\begin{tabular}{c|l|l|l|l}
&Draco\&UMi &Leo IV\&V&LMC\&SMC\\
\hline\hline
$r$ (kpc)&$77\pm 3$&$167\pm 6$&$57\pm 2$\\
\hline
$r_1-r_2$ (kpc)&$2\pm 7$&$24\pm 12$&$13\pm 4$\\
\hline
$[(v^r_2)^2-(v^r_1)^2]\ (10^3\rm{km}^2/\rm{s}^2)$&$2.0$&$3.3$&$4.6$\\
%\hline
%$v_1^t$ (km/s)&$31$&$30$&$21$\\
\hline
\hline
\end{tabular}
\caption{Data relevant for the tangential velocity estimates of 3 candidate binary satellites}
\label{tantab}
\end{table}

If Ursa Minor is closer than Draco, which is marginally preferred by the data, then Eq.~(\ref{dravel}) gives a high tangential velocity for Draco.  Indeed with 1$\sigma$ of confidence $0<r_2-r_1<9$\ kpc and so $(v^t_1)^2$ is greater than 48,000 (km/s)${}^2$, so $v^t_1>220$ km/s.   On the other hand if $0<r_1-r_2<2$\ kpc then $(v^t_1)^2$ will be negative which is clearly impossible, so identical orbits for the two satellites imply that either $r_2>r_1$ and $v^1_t>220$ km/s or else $r_2<r_1-2$\ kpc and $v^1_t<200$ km/s.   Within 1$\sigma$ bounds on the relative radial distances, this tangential velocity is high enough to be measured by Gaia and so this possibility is falsifiable in the near future.

Next we will consider Leo IV and Leo V.  In this case the relative velocities are much greater $v^r_1=13\pm 1$ and $v^r_2=59\pm 3$.  Leo V is receding more quickly than Leo IV.  While they lie upon almost the same line of sight, at $r_2=179\pm 10$ kpc Leo V appears to be more distant than Leo IV at $r_1=155\pm 6$ kpc, so the distance between these satellites is increasing.  In particular the second term in Eq.~(\ref{dravel}) is positive and is between $7,000$ and $20,000$ (km/s)${}^2$.  Again this leads to a large tangential velocity for the Leo's which is easily within the sensitivity of the Gaia mission.

So far we have been unable to present conclusions, only predictions, regarding the scenario in which the elements of each pair have a common progenitor.   The problem is that the tangential velocities of these dwarf spheroidal galaxies are unknown.  However, as mentioned above, the tangential velocities of the Magellanic clouds are well known.  Unfortunately they are so massive that their gravitational interactions cannot be neglected.  Indeed there seem to be gas \citep{maggas} and perhaps stellar \citep{magstel} features created by a collision between the two satellites 200 million years ago.  However a naive application of Eq.~(\ref{dravel}) leads to a tangential velocity for the SMC of $110\pm 30$ km/s, which is more than $3\sigma$ less than the measurement reported by \citet{smc}.  

We conclude that the hypothesis that the satellites in each pair follow similar orbits because they have the same total energy and angular momentum per mass leads to very nontrivial predictions for all three Milky Way pairs.  In the case of two of these pairs the predictions can easily be tested by Gaia.  In the case of the Magellanic clouds this prediction is already strongly excluded by existing data.   Therefore, in what follows we will not impose this hypothesis.

\subsection{Independent motions}

Recall that the radial velocities of the members of each pair agree to within about 30 km/s and their positions agree to within about 30 kpc.  Therefore one may attempt to estimate the differences in their orbits.  Our isotropy assumptions on the relative velocities of the satellites imply 3d relative velocities of order 50 km/s.  On the other hand the Virial theorem, together with their potential energies, lead to total velocities of order 150 km/s in the reference frame of the host.  Therefore one expects a difference in kinetic energy of order 10\%.  In addition, the differences in the radial distances to their hosts leads to a difference in potential energy order of order 10-15\%.  The conclusion of the last subsection suggests that these two differences do not cancel each other, and so we will add them as if they were independent to conclude that the total energies of two elements of a pair differ by 10-20\%.  

As a result the semimajor axes differ by 10-20\% and so the orbital periods differ by of order 15-30\%.   A crude estimate of this already approximate effect is obtained by stating that the separation between the elements of a pair change by 15-30\% of their 150 km/s orbital velocity, leading to a 20-50 km/s or 20-50 kpc/Gyr change in their separation. 

\begin{figure}[!tp]
\begin{center}
\includegraphics[width=0.60\linewidth]{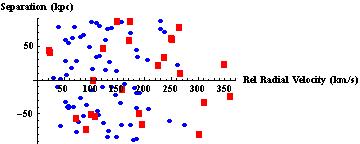}
\caption{Radial velocity differences of Milky Way satellite pairs with radial velocities differing by less than 90 kpc.  The red squares
 are pairs with both members further from the Milky Way than the Magellanic Clouds.} 
\label{mwplotfig}
\end{center}
\end{figure}

What would such a rate of change of the separation of the satellites in each pair imply?  To determine this, one needs to determine the {\it{initial}} separations of the satellite pairs.  This can be extrapolated from the wealth of data on local group satellites assembled by \citet{elenco}.

\noindent
{\bf Milky Way Satellites}

\begin{figure}[!tp]
\begin{center}
\includegraphics[width=0.45\linewidth]{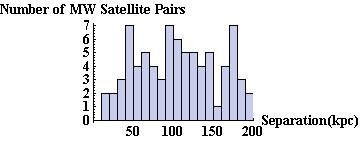}
\includegraphics[width=0.45\linewidth]{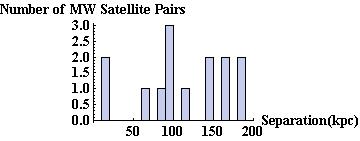}
\caption{Number of Milky Way satellite pairs with radial velocities differing by less than 90 kpc.  {\bf{Left:}} All pairs. {\bf{Right:}} Pairs more distant than the Magellanic clouds.  Note that the number of pairs at small separation does not grow as the square of the separation, as would be the case for a homogeneous distribution.} 
\label{mwtuttifig}
\end{center}
\end{figure}

\begin{figure}[!tp]
\begin{center}
\includegraphics[width=0.6\linewidth]{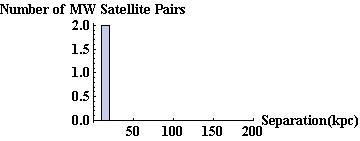}
\caption{Number of Milky Way satellite pairs with radial velocities differing by less than 90 kpc and separation smaller than half of the distance between their midpoint and the Milky Way.  The separations are all less than 30 kpc, suggesting that they formed less than 2 Gyr ago.} 
\label{mwvicinifig}
\end{center}
\end{figure}

The crucial observation is as follows.   There are 26 Milky Way satellites with well-known radial velocities, leading to 325 potential pairs.  Pairs which condensed recently from the same compact progenitor may be expected to have similar line of sight velocities, so we will restrict attention to pairs with line of sight velocities that agree within 90 km/s.  As plotted in Figs.~\ref{mwplotfig} and~\ref{mwtuttifig}, this leaves 95 pairs, although it excludes the Magellanic clouds whose progenitor may have been large.  Now if the distance between the two satellites in the pair is greater than half of the distance from the center of the pair to the Milky Way, then their proximity is well explained by their mutual attraction to the Milky Way and so there is no need to invoke an unobserved common ancestor.  But if we further restrict our attention to satellite pairs separated by at most half of their distance to the Milky Way we find only two pairs, Draco and Ursa Minor and also Leo IV and V, as is shown in Fig.~\ref{mwvicinifig}.   These are separated by just 23 kpc and 25 kpc, although our conditions allowed for separations as large as 90 kpc.   The total volume within a separation of 90 kpc is 60 times larger than that with a separation of 25 kpc, and so a random distribution of pairs would have led to much larger separations.  

There is a natural explanation for the small separations within the common progenitor scenario.  If the size of the common progenitors is of order 30 kpc or less, then one may expect the pairs which condensed from that progenitor to be separated by less than 30 kpc.  

Thus an analysis of the Milky Way satellite pairs seems to suggest that, in the common progenitor scheme, the sizes of the progenitors is at most about 30 kpc.  Now we can return to the crude estimate that the separations are changing by 20-50 kpc/Gyr.  If indeed the initial separations were less than 30 kpc and the separations today are less than 30 kpc, then this gives an upper limit on the time that has elapsed since these satellites condensed of roughly 2 Gyr.  As the pairs are very separated spatially, each seems to have condensed from a different progenitor.  Thus the common progenitor model is fairly constrained, it implies that all 2 or 3 common progenitors in the Milky Way condensed into satellite galaxies in the past 2 Gyr.

The absence of pairs separated by more than 30 kpc and less than half of the distance to the Milky Way has a further implication.  Not only are the pairs which are observed quite young, but there seems to be an absence of older pairs.  If the condensation of progenitors into satellites pairs were common, one would expect pairs of all ages and so of all separations, in conflict with observations.  If on the other hand it were rare, then why would so many events have happened in the past 2 Gyrs?

\noindent
{\bf Andromeda Satellites}

These are all rather strong statements to extract from rather small samples of pairs.  However a similar analysis can in principle be applied to the Andromeda system. One major disadvantage in the case of Andromeda satellites is the comparatively poor knowledge of the distances between the satellites and their host.   Nonetheless the precise knowledge of the angular positions of the satellites yield robust lower bounds on their separations, which in most cases is already sufficient to conclude that the separation between two satellites exceeds half of the distance to Andromeda.

\begin{figure}[!tp]
\begin{center}
\includegraphics[width=0.6\linewidth]{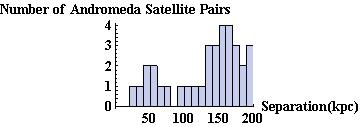}
\caption{Number of Andromeda satellite pairs with radial velocities differing by less than 90 kpc.  Note that it appears to grow more slowly than the square of the distance, with a maximum at separations similar to the distance to Andromeda which is caused by the fact that both elements in each pair are bound to Andromeda.  A homogeneous distribution would have less pairs at small separations.} 
\label{andtuttifig}
\end{center}
\end{figure}

\begin{figure}[!tp]
\begin{center}
\includegraphics[width=0.45\linewidth]{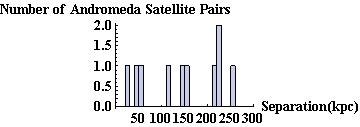}
\includegraphics[width=0.45\linewidth]{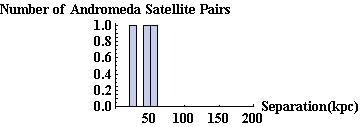}
\caption{Number of Andromeda satellite pairs with radial velocities differing by less than 90 kpc and separated by less than twice the distance between their midpoint and Andromeda.  {\bf{Left:}} The distance between the satellites is set equal to a lower bound on their separation given by assuming that their distances to the Sun are identical.  {\bf{Right:}} The separation between the satellites is a best fit 3-dimensional estimate.  Note that, as in the case of Milky Way satellites, the satellite pairs tend to be at short separations.} 
\label{andvicinifig}
\end{center}
\end{figure}

21 Andromeda satellites have well measured radial velocities, leading to 210 potential pairs.  We will restrict attention to pairs for which the radial velocities with respect to the Sun agree within 90 km/s.  This leaves 75 pairs with separations shown in Fig.~\ref{andtuttifig}.  Now a minimum distance between the pairs can be estimated by fixing the radial distances between each Andromeda satellite and the Sun to be equal, for simplicity we will set them to be equal to the distance to the Andromeda galaxy.  This allows us to further restrict our attention to those pairs separated by a minimum distance which is less than half of their distance to their host Andromeda.    As can be seen in  Fig.~\ref{andvicinifig}, there are 10 such pairs, including the 2 pairs discussed in this note.    The characteristics of these pairs are summarized in Table~\ref{andtab}.

\begin{table}[position specifier]
\centering
\begin{tabular}{c|l|l|l|l|l}
&$d_{\rm{min}}$ (kpc)&$d_{3d}$ (kpc)&$d_{\rm{And}}$ (kpc)&$|v^\odot_1-v^\odot_2|$ (km/s)\\
\hline\hline
And I\&III&$33$&$33^{+12}_{-0}$&$66$&$30\pm 2$\\
\hline
NGC 147\&185&$13$&$59^{+39}_{-36}$&$164$&$11\pm 1$\\
\hline
And XI\&XIV&$56$&$60^{+150}_{-4}$&$133$&$61\pm 5$\\
\hline
And XII\&XIV&$62$&$151^{+170}_{-89}$&$147$&$77\pm 4$\\
\hline
And II\&XIII&$68$&$269^{+45}_{-154}$&$182$&$1\pm 8$\\
\hline
And II\&Triangulum&$61$&$168\pm 28$&$195$&$14\pm 2$\\
\hline
And XVI\&I&$76$&$229\pm 53$&$168$&$9\pm 5$\\
\hline
And XVI\&III&$79$&$233\pm 53$&$177$&$39\pm 5$\\
\hline
And XVI\&XI&$37$&$236^{+65}_{-160}$&$191$&$35\pm 7$\\
\hline
And XVI\&XV&$78$&$122^{+75}_{-38}$&$226$&$46\pm 9$\\
\hline
%$v_1^t$ (km/s)&$31$&$30$&$21$\\
\hline
\end{tabular}
\caption{Andromeda pairs with radial velocities that agree to within 90 km/s and minimum distances that are less than the average distance to Andromeda}
\label{andtab}
\end{table}

Of the 8 potential new pairs, only 1 pair, consisting of Andromeda XI and Andromeda XIV, has a best fit 3d distance (60 kpc) which is less than half the distance to Andromeda.  However these two satellites have a relative radial velocity of 61 km/s, much larger than the others.   For 3 of the other 7 pairs, the relative separation is compatible with less than half of the distance to Andromeda at the $1\sigma$ level, and so associations cannot be ruled out with current data.  %We will now discuss the other seven pairs one at a time. 

In summary, of the satellite pairs in our local group with radial velocities differing by less than 90 km/s and separations of less than half of the distance to their host, both Milky Way satellites have separations of at most 25 kpc.  The situation with Andromeda pairs is somewhat more ambiguous.  Andromeda I and III have a separation which is below 50 kpc.  NGC 147 and 185 have a best fit separation of 59 kpc, but they lie along essentially the same line of sight and a 25 kpc separation is allowed at $1\sigma$.  Several other pairs of Andromeda satellites may be considered, but in each case either the velocities have differences in excess of 50 km/s or else the best fit radial distances lead to large separations.  Thus the distribution of Andromeda pairs is consistent with a low average separation, of order 30 kpc, and no satellites in the 30-90 kpc separation range.  On the other hand, given a homogeneous distribution of satellite distributions in phase space one may have expected pairs with minimum distances in the 40-90 kpc range and relative velocities beneath 50 km/s, but such pairs appear to be missing.

Thus, like pairs of Milky Way satellites, the pairs of Andromeda satellites considered in this note also appear to have condensed less than 3 Gyrs ago and there is mild evidence that older pairs are not present.   This is interesting because, in the case of many of these pairs, although certainly not the Magellanic clouds, star formation would have ceased before the pairs condensed.  This suggests that both the metallicities and stellar populations of the two satellites in a pair should be similar.  For now this is difficult to test in the case of Leo IV and V because of the foreground contamination in Leo V.  However, tangential velocity measurements by the Gaia satellite will help to separate Leo V from the foreground.  On the other hand Ursa Minor is appreciably more metal poor than Draco \citep{metal}, albeit with a difference which is smaller than the metallicity spread.

Just how unlikely such scenarios are depend on the merger history of the Milky Way and Andromeda systems and will be investigated in a future work, however it motivates our search for other potential explanations for the proximities of these gravitationally unbound pairs.

\section{Nongravitational Binding}

Should the common progenitor explanation for the abundance of close pairs of satellites with similar radial velocities be falsified by future data, then what?  There remains another logically consistent explanation.  All conclusions in this article were based on the dynamics of dark matter being governed by Newtonian gravity.  Precision probes establish that general relativity, which in the regime of interest here is well approximated by Newtonian gravity, describes the interactions of baryonic matter extraordinarily well from submillimeter to solar system scales.  Cosmological probes such as the cosmic microwave background (CMB) power spectrum demonstrate that general relativity governs the behavior of dark matter on cosmological scales down to the scales of 10s of kpc at which high wavenumber CMB oscillations were formed prior to recombination.   Information on CMB fluctuations at smaller scales has been erased by Silk damping. 

This program still has one large gap: dark matter below the 10 kpc scale.  If dark matter couples to a particle, besides the graviton, which is lighter than $10^{-25}$eV then it can experience nongravitational long range interactions.  Indeed on the contrary to being excluded by current experimental bounds, such interactions have often been invoked to remedy weaknesses or perceived weaknesses of weakly interacting massive particle (WIMP) models.  Examples involving light scalar fields include for example that of \citet{bec}.  Recently \citet{nobec} have claimed that scalar field models which are capable of reproducing flat rotation curves generically run afoul of the upper bound on the cross sections implied by observations of the Bullet cluster \citep{bullet}.  However this pathology can in turn be cured by the addition of a dark gauge symmetry in models in which dark matter halos are giant 't Hooft-Polyakov monopoles \citep{nani}.  

Whatever the precise structure of these dark force models, there is one common prediction.  Stable dark matter halos in such models may extend beyond their tidal radii, in fact in many models they must.  Cold dark matter WIMPs at astrophysical distances are gravitationally bound and so cannot form stable structures that extend beyond their tidal radii.  Therefore an observation that stable dark matter halos extend beyond their tidal radii would simultaneously falsify all models of dark matter in which the only long distance interaction is gravity, including WIMPs.  The current study certainly does not falsify any models, the common progenitor explanation for the coincidental positions of these pairs is quite plausible and consistent with WIMP phenomenology.  However in the future the outer regions of dwarf spheroidal halos will be mapped both using  lensing and also via tangential velocity measurements of stellar tracers at large radii while their total masses may be determined via measurements of velocity changes of stars in the Milky Way's disk \citep{gaiavel} and so such an exclusion will be feasible.

On the 20th of November the European Space Agency will launch the Gaia space telescope.  Due to its proximity, at 76 kpc, we suggest that the proposed binary system consisting of the Draco and Ursa Minor dwarfs would be a fruitful to observe for three reasons.  First of all, the average of the transverse stellar motions will give a reasonably accurate measurement of the transverse velocities of the two galaxies and so allow a more precise determination of their orbits and thus also their masses.  Second, by determining the transverse velocities of the stars, the degeneracy which plagues the Jeans equation \citep{binney} can be broken, allowing the dispersive motion of the stars to reveal the underlying mass profile.  As these galaxies, unlike the Magellanic clouds, are everywhere dark matter dominated, this will provide a direct measurement of the dark matter halo's shape.  Finally, the transverse velocities can be used to distinguish members of the galaxies from the background.  In Sec.~\ref{maree} we have suggested that the dark matter halos of these galaxies extend far beyond the furthest yet identified stars, which implies that a wealth of members await discovery in the regions where they are outnumbered by nonmembers.  These extra members can be used to trace out, for the first time, the outer regions of a dwarf spheroidal galaxy's dark matter halo.

\subsection*{Acknowledgments}

I would like to thank  Malcolm Fairbairn and Sven Bjarke Gudnason for discussions and comments on this draft.  JE is supported by the Chinese Academy of Sciences Fellowship for
Young International Scientists grant number 2010Y2JA01.

\end{document}